\documentclass[sigconf]{acmart}

\usepackage{balance} 

\usepackage{booktabs} 
\usepackage{graphicx}
\usepackage{amsmath} 
\usepackage{amssymb}  
\usepackage{algorithm}
\usepackage{algpseudocode}
\usepackage{url}
\usepackage[nolist,nohyperlinks]{acronym}
\newcommand{\squeezeup}{\vspace{-2.5mm}}

\copyrightyear{2019} 
\acmYear{2019} 
\setcopyright{rightsretained} 
\acmConference[ICCPS '19]{10th ACM/IEEE International Conference on Cyber-Physical Systems (with CPS-IoT Week 2019)}{April 16--18, 2019}{Montreal, QC, Canada}
\acmBooktitle{10th ACM/IEEE International Conference on Cyber-Physical Systems (with CPS-IoT Week 2019) (ICCPS '19), April 16--18, 2019, Montreal, QC, Canada}
\acmDOI{10.1145/3302509.3313323}
\acmISBN{978-1-4503-6285-6/19/04}

\acmPrice{15.00}

\begin{document}

\title[Contract-based Hierarchical Resilience Framework for CPS]{Demo Abstract: Contract-based Hierarchical Resilience Framework for Cyber-Physical Systems}


\author{Daniel Jun Xian Ng}
\affiliation{Nanyang Technological University, \\
	50 Nanyang Avenue, Singapore}
\email{danielngjj@ntu.edu.sg}

\author{Arvind Easwaran}
\affiliation{Nanyang Technological University, \\
	50 Nanyang Avenue, Singapore}
\email{arvinde@ntu.edu.sg}

\author{Sidharta Andalam}
\affiliation{Delta Electronics, \\
	50 Nanyang Avenue, Singapore}
\email{sidharta.andalam@deltaww.com}


\begin{abstract}
	
	This demonstration presents a framework for building a resilient Cyber-Physical Systems (CPS) cyber-infrastructure through the use of hierarchical parametric assume-guarantee contracts. A Fischertechnik Sorting Line with Color Detection training model is used to showcase our framework. 
	
\end{abstract}

%
%



\maketitle

\section{Introduction}
Industry 4.0~\cite{iFour} has garnered much interest in the manufacturing industry to create smart factories. This move towards smart factories requires incorporating more computational devices for decentralized decision-making and more sensors on the factory floors. The availability of more data provides better transparency in making the appropriate decisions during runtime and for fault recovery. With all of these devices interconnected, a robust networking infrastructure becomes crucial for system monitoring and ensuring the availability and timely arrival of priority packets.

Disruptions to the above cyber-infrastructure due to faults will be of severe consequence and thus there is a need for a resilient infrastructure. As these manufacturing systems become increasingly complex with distributed infrastructure, it also becomes harder to develop and maintain large amounts of application as well as fault handling code.

\begin{figure}[htbp]
	\centering
	\includegraphics[scale=0.25]{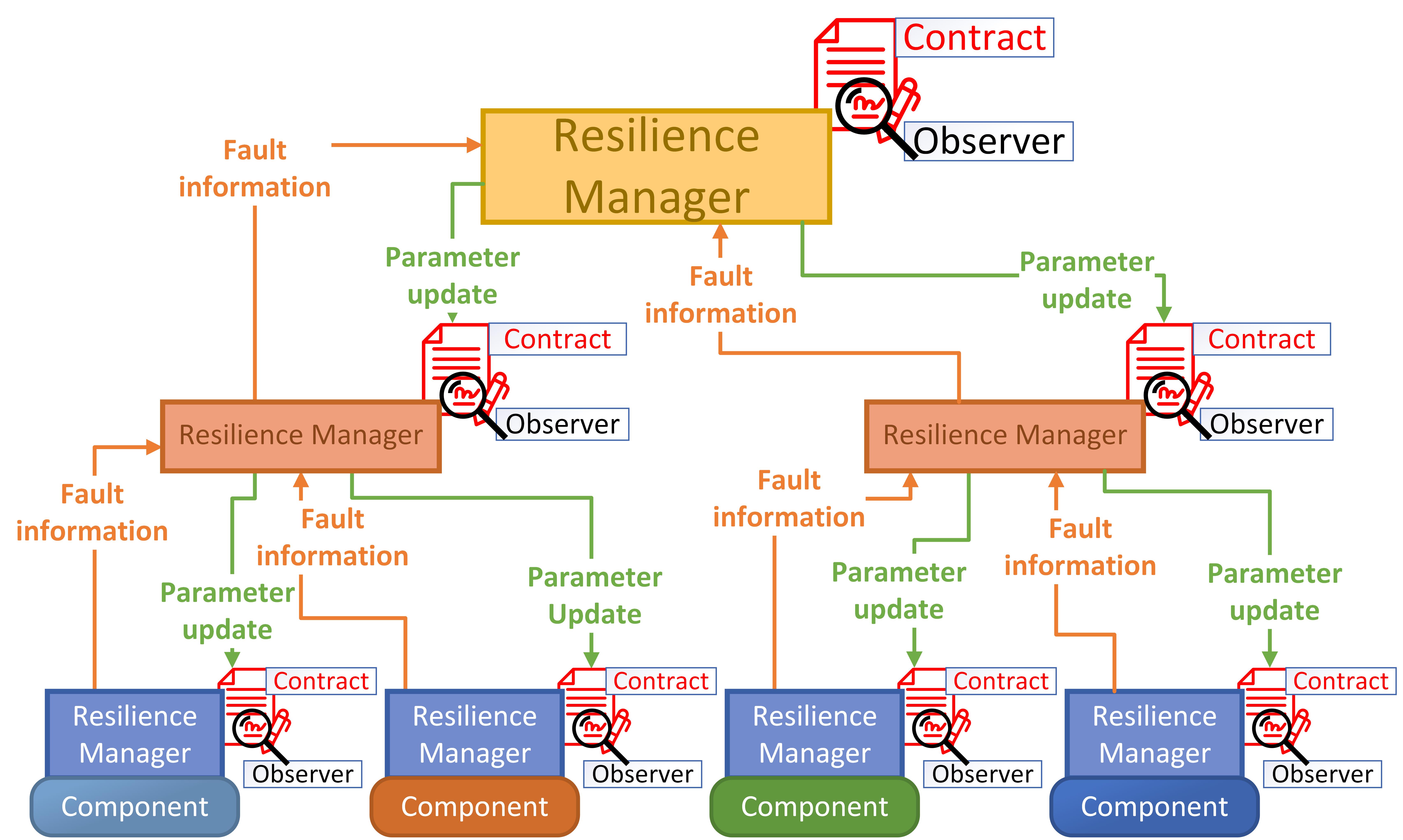}
	\caption{Hierarchical Contract-based Resilience Framework}\label{fig:hierFrame}
	\includegraphics[scale=0.25]{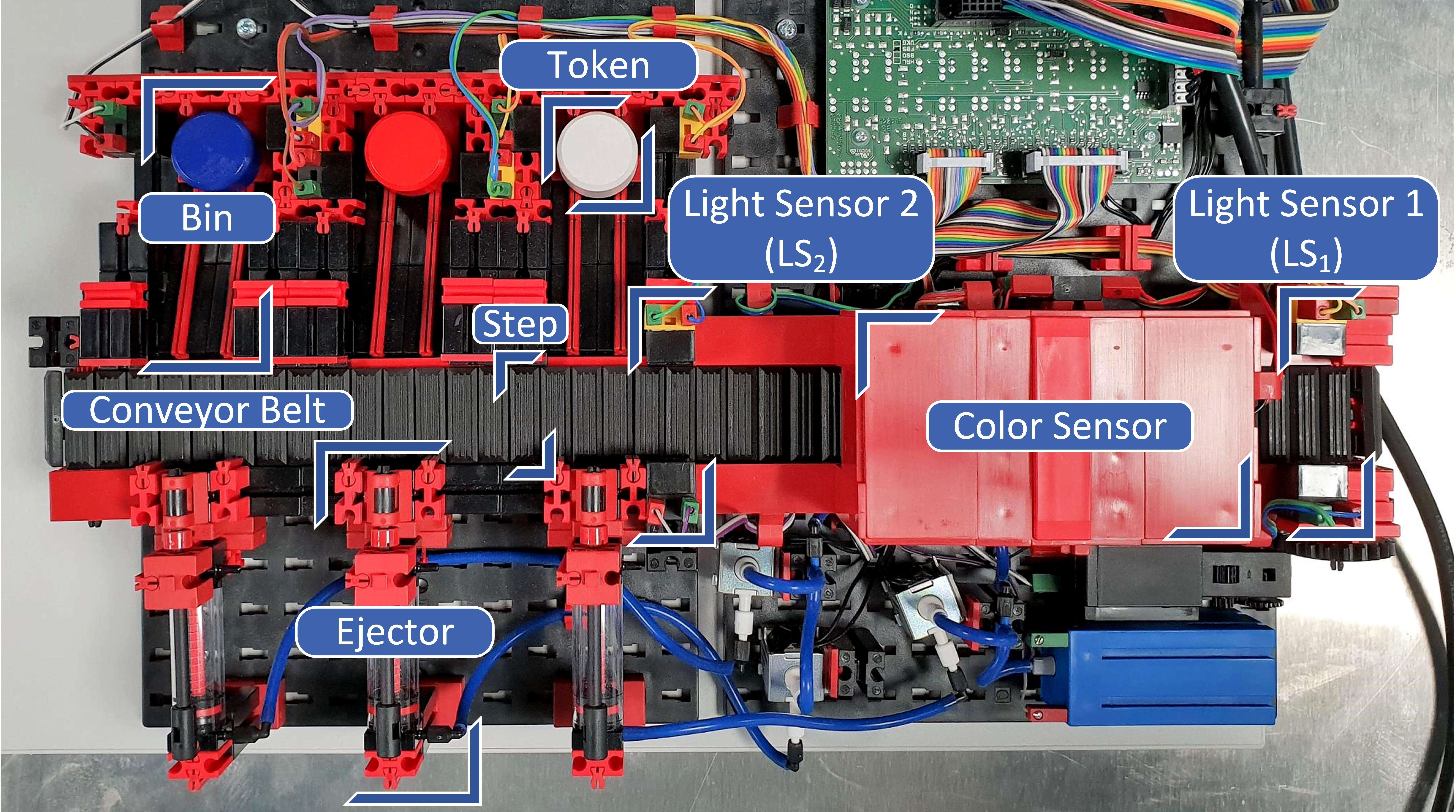}
	\caption{Fischertechnik Sorting Line with Colour Detection}\label{fig:testbed}
\end{figure}

To overcome these problems, we propose our Contract-based Hierarchical Resilience Framework for Cyber-Physical Systems as shown in Figure~\ref{fig:hierFrame}. Our framework consists of system \textit{components} (e.g. sensors, actuators and controllers), \textit{Resilience Managers} (RM) and \textit{observers}~\cite{ISORC_CLAIR}. We use assume-guarantee contracts~\cite{AG_Contracts} to capture the guarantees provided by system components (i.e., requirements) which are monitored by observers during runtime. Deviations from these guarantees trigger a fault by the observers and this is reported to the RM associated to it. A set of contracts is managed by an RM in the framework and the RM decides on the recovery response. The RMs and contracts are also structured in a hierarchy and we use parametric assume-guarantee contracts~\cite{P_Contract} to allow for scalability and to reduce communication overheads between RMs~\cite{COMP_HContract}. The recovery response depends on the combination and extent of contract violations; an RM may either respond by changing contract parameters (i.e., modify and hence potentially degrade component performance) or propagating the fault to a higher level RM. With a hierarchy, we can decompose contracts into sub-contracts which allow for independent lower-level decision-making by the RMs. This hierarchy also enforces a strict coordination protocol among the RMs when recovery solutions cannot be found at lower levels. Further details of this framework can be obtained from related publications~\cite{ISORC_CLAIR,COMP_HContract}. 

\section{Demonstrator}
We illustrate our resilience framework on a Fischertechnik Sorting Line with Color Detection training model as shown in Figure~\ref{fig:testbed}. 

\begin{figure}[htbp]
	\centering
	\includegraphics[width=\columnwidth]{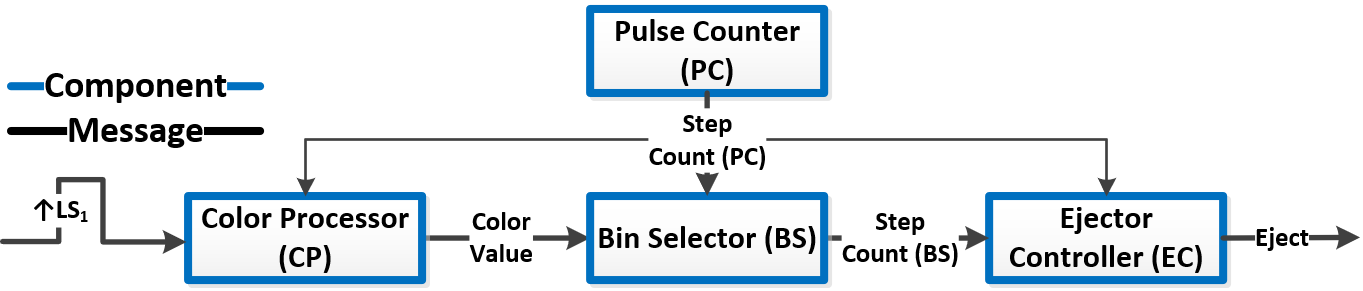}
	\caption{Operation flow}\label{fig:OpFlow}
\end{figure}

It has two light sensors ($ LS_1 $, $ LS_2 $), a color sensor, a conveyor belt, three ejectors and three bins for token storage. A motor controller (MC) regulates the belt's rotation and a pulse counter (PC) tracks the belt's steps. Tokens are placed on the conveyor belt at $ LS_1 $ and it goes through the color sensor, triggered by the color processor (CP). A decision-making component, a bin selector (BS), determines the color of the token and sends the information to the ejector controller (EC). The EC then determines when to eject the token into its designated bin.  This inter-component dependency creates an end-to-end latency requirement from the start where $ LS_1 $ is located, to the end where the bin resides. The operation flow of the testbed and its latency requirement are illustrated in Figure~\ref{fig:OpFlow}. 

A fault could lead to a longer computation time of a component, violating its latency contract. As a result, the end-to-end latency requirement may no longer be satisfied. In our case study in~\cite{ISORC_CLAIR}, the ejector failed to push the token into the designated bin as the PC component was unable to meet its latency requirement of 10ms, resulting in a delay of providing an accurate step reading of the conveyor belt to EC. The resilience manager's recovery response was to change the behavior of PC to shorten its response time, rectifying the fault.  In~\cite{COMP_HContract}, components CP, BS or EC each have a latency contract to guarantee their response times. When faults occur, leading to longer computation times, our resilience framework could rectify this problem by adjusting multiple contracts' latency parameters at runtime. This ensures that the end-to-end requirement is once again satisfied. In this scenario, the higher level RM may choose to reduce the conveyor belt's speed to satisfy the end-to-end timing requirement, whenever the underlying fault is significant. However, because of the flexibility offered by the contract hierarchy, it is also possible that this RM is able to compensate for a timing fault in one component using slack from another, thus avoiding this degradation in some cases.

\begin{figure}[htbp]
	\centering
	\includegraphics[scale=0.25]{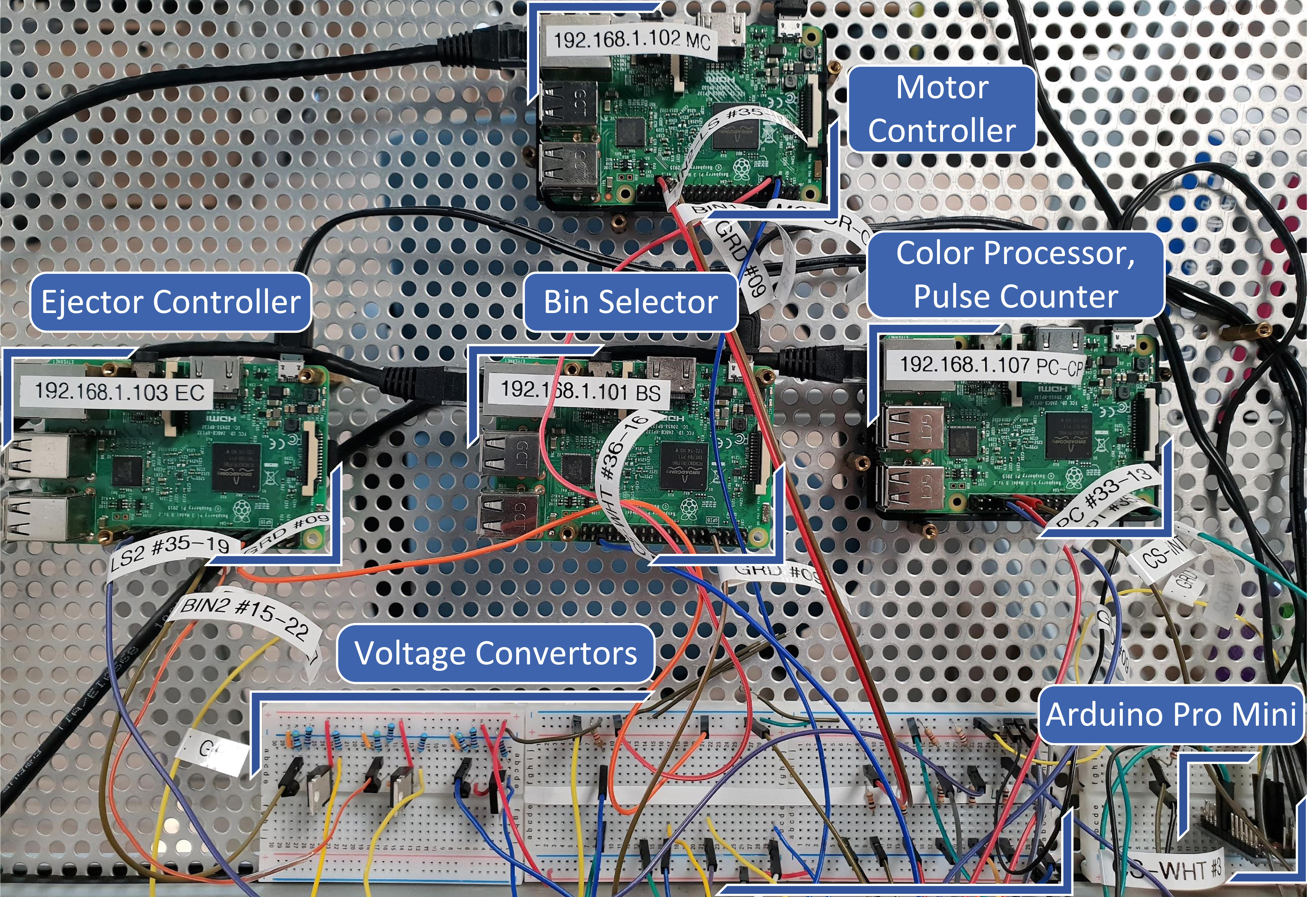}
	\caption{Computation Nodes}\label{fig:compute}
\end{figure}

\begin{figure}[htbp]
	\centering
	\includegraphics[width=\columnwidth]{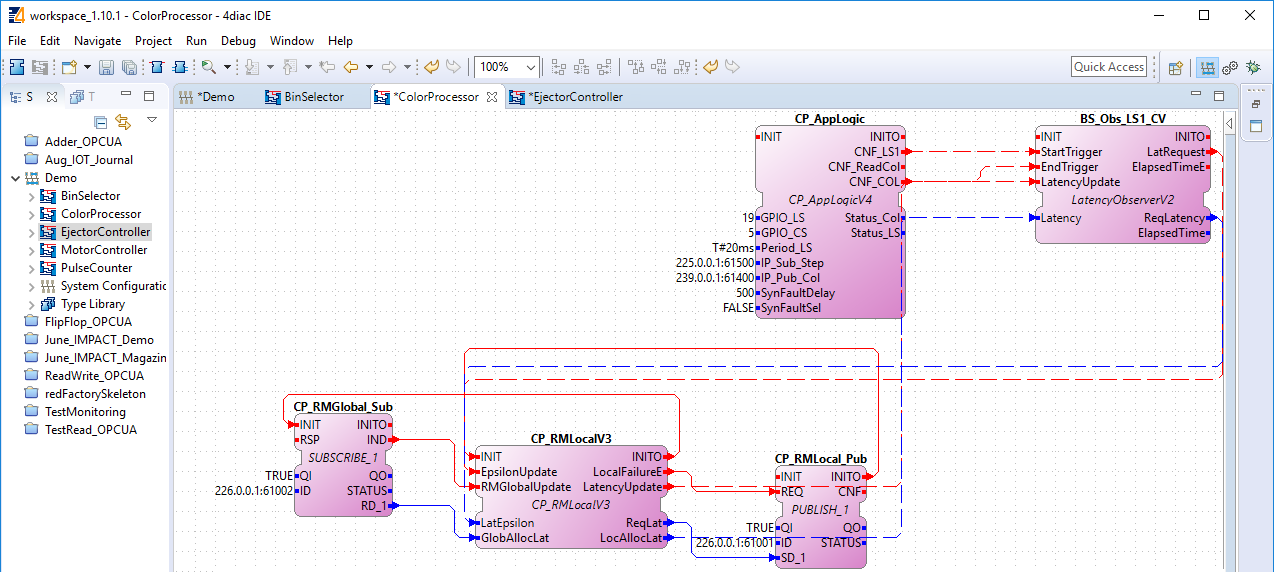}
	\caption{4DIAC Integrated Development Environment}\label{fig:4diac}
\end{figure}

The testbed is coupled with four Raspberry Pi 3s (RPIs) shown in Figure~\ref{fig:compute} which serve as the computational devices for the five components mentioned. The software implementation of the resilience framework, as well as the sorting line application is done on 4DIAC~\cite{4DIAC}, an open source framework for industrial automation and control that follows the IEC 61499 standard~\cite{Zoitl}. It provides a development environment shown in Figure~\ref{fig:4diac} which has the function blocks for the CP component as well as a runtime environment \textit{FORTE} which runs on the RPIs. The lower three pink function blocks shown belongs to the Resilience Manager; the top left block represents the application logic and the rightmost block shows the observer. This arrangement allows for segregation between application code and fault handling code, thus enabling systematic development. Communication between function blocks is handled through the use of an in-built Publisher/Subscriber mechanism where the RPIs are interconnected using an Ethernet network switch.

\squeezeup

\begin{acks}
	This research work under project SLI-RP4 was conducted within the Delta-NTU Corporate Lab for Cyber-Physical Systems with funding support from Delta Electronics Int'l (Singapore) Pte. Ltd. and the National Research Foundation (NRF) Singapore under the CorpLab@University Scheme.
\end{acks}

\bibliographystyle{ACM-Reference-Format}
\bibliography{Bibliography}

\end{document}